# *Continuous Wide-Field Optical Monitoring for Very Early-Phase Transient Discovery*

A white paper in response to the ESO Call about the Expanding Horizons program by:

Massimo Della Valle[1], Maria Teresa Botticella[1], Enrico Cappellaro[2], Roberto Ragazzoni[2,9], Matteo Aliverti[3], Carmelo Arcidiacono[2], Lorenzo Amati[8], Andrea Baruffolo[2], Maria Grazia Bernardini[3], Giovanni Boato[9], Fabrizio Bocchino[4], Francesco Borsa[3], Mohamed Yahia Bournane[2,9], Enzo Brocato[5], Giovanni Bruno[6], Paolo D'Avanzo[3], Nancy Elias-Rosa[2], Silvio Di Rosa[2], Diego Farias[1], Jacopo Farinato[2], Davide Greggio[2], Adriano Ingallinera[6], Luca Izzo[1], Marco Limongi[5], Demetrio Magrin[2],  Marco Marongiu[7], Andrea Melandri[5], Giusi Micela[4], Matteo Murgia[7], Valerio Nascimbeni[2], Salvatore Orlando[4], Antonino Petralia[4], Vincenzo Petrecca[1], Maura Pilia[6], Silvia Piranomonte[5] Andrea Possenti[7], Kalyan Radhakrishnan[2], Oleksandra Rebrysh[2], Simone Riggi[6], Irene Salmaso[1], Giovanni Scandariato[6], Corrado Trigilio[6], Simone Zaggia[2]

1: INAF – Osservatorio Astronomico di Capodimonte, Napoli; 2: INAF – Osservatorio Astronomico di Padova; 3: INAF – Osservatorio Astronomico di Brera, Milano; 4: INAF – Osservatorio Astronomico di Palermo; 5: INAF – Osservatorio Astronomico di Roma; 6: INAF – Osservatorio Astronomico di Catania; 7: INAF – Osservatorio Astronomico di Cagliari; 8: INAF – Osservatorio di astrofisica e scienza dello spazio; 9: Dip.to di Fisica e Astronomia, Università di Padova

**Abstract.** The study of transient phenomena in a multimessenger context is expected to remain a major pillar of astrophysical discovery in the decades ahead. Supernovae, Kilonovae, Black-Hole formation, Novae, GRBs, and tidal disruption events are prime examples, as their earliest phases link electromagnetic radiation to gravitational waves, neutrinos, and high-energy emission. Yet, the critical physics connecting these messengers unfolds within minutes to hours, while traditional surveys revisit the same region of the sky on the scale of days/weeks, missing the moment when the event actually begins. Current survey facilities excel at answering what happened (*classification*) and how often (*rate*)**,** but essentially fail in addressing how it happened, how it begins, and how it couples to gravitational waves, neutrinos, or high-energy emission**.** Continuous wide-area optical monitoring, as proposed in this document, removes this limitation. The traditional approach, where a GW or neutrino alert triggers electromagnetic follow-up, is now complemented, and sometimes reversed: early electromagnetic discoveries can prompt searches for weaker gravitational waves or neutrino signals that would otherwise be missed. In the Einstein Telescope era, wide-field optical monitoring will allow us to quickly find the optical counterparts of gravitational-wave events and understand their physics. At the same time, a telescope capable of continuous monitoring provides immediate scientific value for planetary defense, space-debris tracking, stellar variability, exoplanets transit monitoring, accretion-driven activity, with one more important item: when we step into a new observational territory, the *true* discoveries are often the ones we did not expect. In this vision, continuous time-domain astronomy does not replace classical surveys: it completes them by supplying the missing temporal dimension. Follow-up observations remain essential, but they now begin at the physical onset of the event rather than after its evolution is already underway.

**1. Introduction - From Exceptional Omens to a Continuous Universe.** For most of human history, the changing sky was an exception. Civilizations carefully recorded the rare appearance of "guest stars," from the detailed notes of Chinese court astronomers to the first discussions of "new stars" by European astronomers of the Renaissance. A single bright supernova (SN) roughly once per century, was enough to leave an imprint in culture, religion, and early science. What was once regarded as extraordinary is now recognized as "normal". In the observable Universe out to redshift $z \cong 4$, roughly five supernovae explode every second (Ref. 1) and several gamma-ray bursts (GRBs) occur each day. The transient sky is not rare: it is continuous. And yet, despite the frequency of these events, our ability to study them remains rather limited. They are unpredictable in time and position, and the current architecture of ground-and space-based telescopes relies overwhelmingly on follow-up observations triggered after discovery. Most of what we know comes from the middle or late stages of the explosion. The earliest phases, where the physical conditions are most extreme and most informative remain largely unseen. Time-domain astrophysics is now a central pillar of modern astronomy. A large fraction of high-energy astrophysics is encoded in transient phenomena, which trace the endpoints of stellar evolution, compact-object mergers, collisions, and accretion instabilities. Among the most scientifically valuable are SNe, GRBs, Novae, and Kilonovae (KNe) associated with neutron-star mergers. Each of these classes of astrophysical objects probes fundamental physics: i) Core-collapse SNe test the nuclear equation of state and generate neutrino bursts and can provide measurable Gravitational Waves (GWs) emissions; ii) Type Ia SNe originate from white-dwarf binary evolution are standard candles on which the modern cosmology characterized by the discovery of the accelerated expansion of the universe is based (Ref. 2, 3 and 4); iii) GRBs trace highly relativistic jets, the star formation at very high redshift and the formation of stellar black holes or magnetars; iv) Novae probe thermonuclear ignition and mass transfer in compact binaries and possibly are the most effective Lithium factory of the Universe; v) Kilonovae encode r-process nucleosynthesis, linking heavy-element production to gravitational-wave sources.

**2. The Observational Bottleneck: We Always Arrive Too Late.** The scientific potential of these transients lies primarily in their very early emission. The core collapse of a SN occurs on extremely short timescales, of the order of 0.1s-1s and is followed by a brief burst of neutrinos lasting only a few seconds. The subsequent shock breakout of a core-collapse SN lasts dozen minutes to a few hours. The rising thermal component and cocoon emission connected to core-collapse SNe with long GRBs or neutron-star merger last from minutes to less than one day (Ref. 5). The UV/optical flash of circumstellar interaction in SNe-Ia deflagration lasts at the most some hours. The first photometric signatures of fast novae eruptions evolve on timescales of hours, not days. In the multimessenger evolution of GW170817, the very early phase is driven almost entirely by gravitational-wave emission, which lasts only a few tens of seconds and marks the true beginning of the event. Detecting the optical rise of such a transient with a classical survey telescope would therefore depend on a rare stroke of luck, the instrument would have to be observing exactly the right region of the sky at precisely the right moment, within a time window of seconds/minutes, to capture the onset of the electromagnetic emission. One of the central motivations behind the concept of *continuous monitoring* is to transform very early-phase detection from exceptional unachievable coincidence into an expected outcome. Even with state-of-the-art surveys, early discovery remains difficult because their early-time timescales are shorter than any observational cadences achievable today. For example, the next-generation Vera Rubin Observatory LSST will survey the sky with a typical cadence of ~ 3 days per field. In other words, systems whose early evolution unfolds on minutes-to-hours timescales are observed with a day-scale cadence. As a result, they are missed at their physical origin and detected only after they have decoupled from their multi-messenger phase (i.e. neutrinos, gravitational waves, prompt gamma rays). The current surveys answer *what* happened and *how often* (e.g. Ref. 6) but not *how* it physically happened.

**3. The Missing Dimension in Time-Domain Astronomy**. The breakthrough we propose is not a matter of instrumental size, but of temporal sampling. Rather than waiting for the Universe to reveal a transient during a scheduled visit, a continuous monitoring could be designed to be watching when it happens. If the relevant physics unfolds within ≲3 hours and the sky is revisited on average by LSST every ~72 hours, the chance of sampling the initial evolution, *assuming the transient happens exactly in the LSST field being observed*, is ~4%. However, LSST observes only ~10 deg² (Ref. 7) at a time over a survey footprint of ~18,000 deg², reducing the practical probability of catching the early optical rise to ~0.002% (that is 1 in 50,000). In other words, early detection is not just unlikely, it is statistically negligible under non-continuous survey strategies. With similar considerations, assuming continuous monitoring over ~10,000 deg², a cadence of 3 hours yields a near-unit detection probability for transients evolving on a "hour" timescale. Pushing the cadence to 5 minute dramatically improves the situation: phenomena evolving on timescales up to 3 hours are again detected with near certainty, while even extremely fast transients with ~10 s timescales are sampled with a probability of ~ 1 in 30. This removes the dependence on knowing in advance where the event will occur and transforms early-time detection from a rare coincidence into an operational capability. Although a 5-minute exposure, with a 2-4m class *MezzoCielo-like* (Ref. 8) telescope, allows the survey to reach limiting magnitudes considerably fainter than m = 21, we conservatively adopt m = 21, as the effective limit in order to ensure adequate S/N ratio for tracking the early temporal evolution of the transient, as soon the source fades. A survey reaching m = 21 can therefore detect CC-SNe with M = −17, out to ≈400 Mpc, corresponding to a sampled volume of ≈2.6 × 10$^8$ Mpc³. For a volumetric SN rate of ≈10$^{-4}$ yr$^{-1}$ Mpc$^{-3}$ (Ref. 9), this implies ~ 2.6 × 10$^4$ SNe per year over the full sky, which scales to about 6500 events per year over a monitored area of 10,000 deg². With a sampling probability of ~ 1/30 for 10-second timescale phenomena, an annual yield of ~6,500 CC-SNe implies about 200 events per year sampled during their earliest seconds, enabling statistical and physical studies of the very early emission. With similar arguments applied to the case of KNe, assuming an absolute magnitude of M = −16 for GW170817, the survey probes distances out to ≈250 Mpc and for a volumetric rate up to ≅ 1200 Gpc$^{-3}$yr$^{-1}$ (Ref. 10) this yields ≅ 80 KNe per year over the full sky, or ≅ 20 events per year over 10,000 deg², implying an expected rate of a few events per year sampled during the very early stages.

**4. Scientific Returns.** Early-time optical access probes a broad and previously inaccessible regime of transient astrophysics. Within the first minutes to hours of an explosive event, the dominant physical processes, energy release, shock formation, mass ejection, jet launching, or even the absence of any visible breakout, unfold on timescales far shorter than those probed by classical survey cadences. In CC-SNe, the shock breakout and early cooling emission reveal the progenitor's radius, envelope structure, and explosion energy, before later diffusion phases wash out this information. In KNe, the cocoon emission and early thermal component trace the geometry and composition of the merger ejecta and place constraints on r-process production at its origin. GRBs reveal their engines in real time through the onset of relativistic jets and cocoon expansion (Ref. 5) and the dependence of mass-loss on accretion history. Equally important are the transients defined by what fails to appear: direct collapse to a black hole ("failed supernovae") may produce no classical optical explosion; instead, the electromagnetic signature is the rapid disappearance of the progenitor and a faint or absent breakout. Detecting such events requires continuous monitoring, not follow-up. The same holds for "silent" gravitational-wave ejections or precursor flares preceding compact-object mergers, where the decisive information takes time windows measured in minutes. Early-time access therefore transforms the achievable science: it allows optical counterparts of neutrino or GW triggers to be identified while the physics is occurring. Phenomena once considered observationally out of reach become part of the accessible parameter space of routine time-domain astronomy.

**5. Continuous monitoring turns electromagnetic discovery into the first step of multi-messenger astronomy, rather than the last.** Traditionally, the field has operated in a reactive mode, with GWs or neutrino alerts triggering optical searches after the earliest and most informative phases have already passed. With continuous wide-area monitoring, early electromagnetic signatures can be detected independently of external triggers, enabling targeted retrospective searches in GW and neutrino data, even for sub-threshold or poorly localized events. As estimated in Section 3, this approach implies that of order ~20 kilonovae per year can be detected electromagnetically over 10,000 deg², providing a substantial and well-defined sample for gravitational-wave and neutrino follow-up. This capability is particularly relevant in the Einstein Telescope era, where many detections will be marginal or weakly localized. In this framework, electromagnetic observations do not simply respond to multi-messenger alerts: they help generate them, shifting the paradigm from unidirectional follow-up to bidirectional discovery. The "light" can point back to the archives of LIGO/Virgo/KAGRA or km$^3$/IceCube/SuperKamiokande and reveal signals that would never have triggered an alert autonomously.

**6. By-Products.** Although every early-time transient astrophysics is the primary driver of continuous wide-field optical monitoring, such a facility also delivers high-value scientific by-products. Continuous sky coverage supports planetary defense through near-Earth object detection and space-debris monitoring, while enabling systematic studies of stellar variability, systematic studies of exoplanet's transits, late-stage stellar evolution, and accretion-driven activity in AGNs and Quasars. Finally, expanding observational parameter space has historically led to unexpected discoveries: from the CMB, to pulsars and cosmic acceleration, then demonstrating that new ways of observing the sky reveal physical processes we did not anticipate.

**7. Conclusions.** The fundamental astrophysical question addressed by *Continuous Very Early-time Monitoring* is how energy, momentum, and freshly synthesized elements are injected into the Universe by catastrophic events. This injection regulates stellar feedback, chemical enrichment, compact-object formation, and ultimately galaxy evolution. These processes are not encoded in late-time light curves or spectra, but are set during the first seconds to hours of the event, when shocks form, jets are launched, neutrino and gravitational-wave emission peaks, and the geometry and composition of the ejecta are established, before the different messengers decouple and trace the different stages of the events. Current surveys, with day-scale cadences miss this phase, sampling only the aftermath rather than the physical origin of the phenomenon. Continuous optical monitoring adds the missing temporal layer to multi-messenger astronomy, transforming very rare coincidences into routine discoveries, offering not snapshots of a dynamic Universe, but its unfolding. Not a single frame, but a story.

**Technology developments & data handling requirements.** We briefly outline how the scientific goal of a continuous wide field monitoring could be achieved. In principle, this can be arranged using replicas of existing telescopes like Schmidt or FlyEye designs (Ref. 11) that have proven to reach equivalent Field of View of the order of 100 square degrees. As from a single point on the globe the sky above 30 degrees of elevation (airmass two) corresponds to a quarter of the whole celestial sphere, amounting to $10^4$ square degrees, the typical facility consists of about one hundred of such telescopes. Apertures ranging in the 2 to 4m class would requires the exploitation of the FlyEye technology at such aperture although there are not obvious showstopper other than the complexity of operating such a large facility. The extension of monocentric optical systems to large apertures, ~4m, has been proposed using a tiled spherical device filled with low refractive index, high transparency industrial fluid. This concept has been proposed and nicknamed "MezzoCielo" (Ref. 8). Several studies and prototypes, including one as technological demonstrator in support of the Sardinia's site for the Einstein Telescope, are being carried out. Whenever technical evidence suggests a practical limit in the aperture of the monocentric sphere a small array (like a 2x2 or 3x3) of these could be considered in order to reach the desired equivalent aperture. Of course, if one wants to exploit continuously a full whole sky monitoring regardless the day and night limitations would requires the construction of several (4 to 8) of these facilities in different longitudes and in the two latitudes unless one limits to a success rate of 50% and to the southern hemisphere, where just one of the facility described above is reasonably enough. Generally speaking a total number of approx. less than 1000 camera are envisioned. This is consistent with projects for full sky coverage with much smaller aperture (like the 20cm equivalent aperture of ARGUS, Ref. 12). At this scale, the system rapidly crosses the petapixel threshold, then the main technological challenge lies furthermore in handling and analyzing the massive data volumes produced, making advanced data processing and AI-based techniques essential.

**Ref. 11:** Arcidiacono et a. (2024) SPIE proc. 13096, 13096A7; **Ref. 12:** Law, et al., (2022) PASP 134, 1033, 035003